# An Electromagnetic-Flux-Distribution Model for the Analyses of Superconducting Josephson Junction Circuits and Quantum Phase-Slip Junction Circuits

Yongliang Wang

***Abstract*—Josephson junctions and the quantum phase-slip (QPS) junctions are two quantum circuit elements introduced by superconducting electronics to create various hybrid circuits. Josephson junctions bring the developments of superconducting quantum interference devices (SQUIDs) and single-flux quantum (SFQ) digital circuits; as the dual element of Josephson junctions, QPS junctions are used to design the new devices dual to Josephson junction circuits. In order to bridge the gap between superconducting and non-superconducting circuits, this article presents an electromagnetic-flux-distribution model to unify the superconducting and non-superconducting circuit analyses. This model redefines the circuit laws and functions of circuit elements using the conventional electric variables; it provides the unified circuit equations to depict the working principles of circuits viewed from electric and magnetic fields; it also derives the mathematical expression of duality principles between Josephson junction and QPS Junction circuits. The application of this electromagnetic-flux-distribution model is demonstrated in the analyses of a Josephson junction circuit and its dual QPS junction circuit. It shows that Josephson junction circuits are the magnetic-flux-distribution systems, while QPS junction circuits are the electric-flux-distribution systems, they are modulated by the Josephson effect and the phase-slip effect.**

## I. Introduction

THE Josephson junction integrated circuits, such as the rf and dc superconducting quantum interference devices (SQUIDs) [1], have been developed from the simple single-loop devices into the multi-loop integrated circuits with hundreds and thousands of junctions, such as SQUID array [2], superconducting quantum interference filter (SQIF) [3], and single flux quantum (SFQ) logic circuits [4]. They are the complementary devices and circuits to the semiconducting electronics in both analog and digital domains [5].

The Quantum phase-slip (QPS) junction [6] is another type of nonlinear superconducting element, which is one of the dual elements of the Josephson junction. Its charge-based quantum physics dual to Josephson effects have been investigated, and the dual circuit of dc SQUID with two QPS junctions has been demonstrated [7, 8].

Josephson junction and QPS junction circuits will be integrated with the field-effect-transistors (FETs) to develop the superconducting-semiconducting hybrid circuits. For example, Fig. 1 shows a hybrid analog circuit, in which a dc SQUID is amplified by a field-effect-transistor (FET); in the superconducting-semiconducting hybrid digital system, SFQ digital circuits are integrated with the memory elements based on the complementary metal-oxide semiconductors [9].

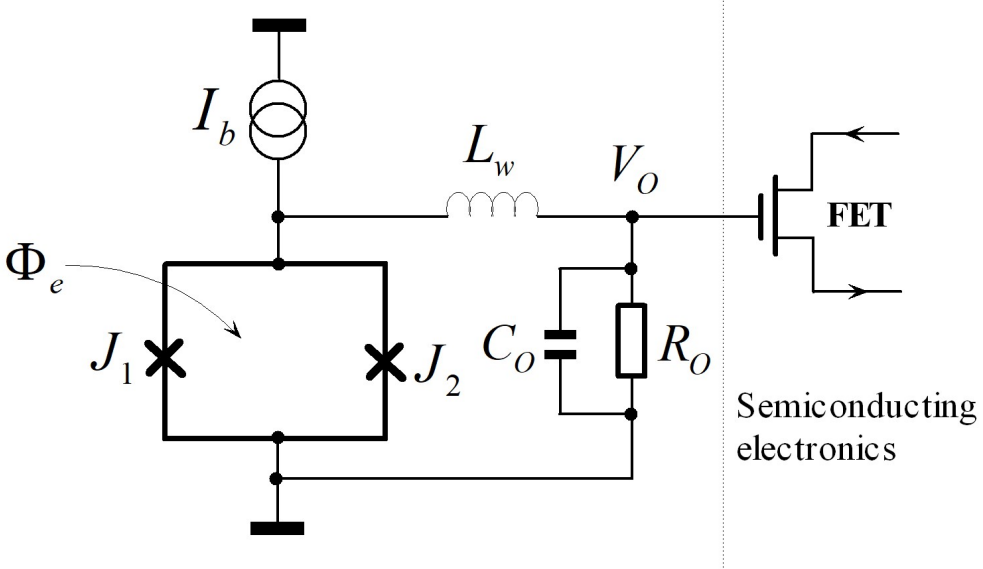

Fig. 1. A superconducting-semiconducting hybrid circuit consists of a dc SQUID and a field-effective transistor.

There are three inconveniences in the analyses of Josephson junction and QPS junction hybrid circuits. First, the variables for the superconducting and non-superconducting circuit elements are not unified. The function of Josephson junction and circuit laws are defined with the phases of the macroscopic wave-functions [11], but the conventional circuit elements and the Kirchhoff's laws [12] use the current and voltage as variables. Second, the circuit equation of superconducting circuits is unavailable to users in the simulation tools; with the absence of circuit equations, it is difficult to study the dynamic working principle inside the Josephson junction and QPS junction circuits. Finally, there is a lack of mathematic expression of the duality principle [13, 14] between Josephson junction and QPS junction circuits. The mathematically defined duality principle is used to find the equivalent Josephson junction circuit for any QPS junction circuit precisely, with which the QPS junction circuit can be simulated using the simulation tools of Josephson junction circuits.

Yongliang Wang is with the State Key Laboratory of Functional Materials for Informatics, Shanghai Institute of Microsystem and Information Technology, Chinese Academy of Sciences and the Center for Excellence in Superconducting Electronics (CENSE), Shanghai 200050, China (e-mail: wangyl@mail.sim.ac.cn).

This article introduces an electromagnetic-flux-distribution model to unify the circuit equations and dynamic models of the superconducting Josephson junction and QPS junction circuits. Based on this model, the Josephson junction circuit is treated as the magnetic-flux distribution system, in which, the magnetic fluxes are distributed inside the inductively coupled loops, and the non-inductive branches with Josephson junction work as the magnetic-flux pumps; the QPS junction circuit is treated as the electric-flux distribution system, in which the electric fluxes are distributed in the capacitively coupled nodes, and the non-capacitive branches with QPS junction work as the electric-flux pumps. From the circuit equations, the mathematical expression of the duality principle between the Josephson junction and QPS junction circuits is derived. This electromagnetic-flux-distribution model is derived using the conventional circuit variables and laws; it bridges the gap between the superconducting and non-superconducting circuits.

## II. Theory

In this section, we firstly introduce the concept of the electromagnetic-flux to rewrite the circuit laws and the functions of circuit branches; then we introduce the electromagnetic-flux-distribution model to derive the unified circuit equations for both Josephson junction and QPS junction circuits. Finally, from the unified circuit equations, we derive the mathematic expression of the duality principle between Josephson junction and QPS junction circuits.

### A. Electric Fluxes in Nodes and Magnetic Fluxes in Loops

Fig. 2 depicts all four electric variables in the analysis of conventional electric circuits, i.e., current ($i$), voltage ($v$), magnetic flux ($\Phi$), and charge ($Q$). They are associated with three basic linear elements, i.e., resistor ($R$), inductor ($L$), and capacitor ($C$). The voltage is the derivative of the magnetic flux, while the current is the derivative of the electric charge, with respect to time. Therefore, the magnetic flux and electric charge are two basic physical quantities in electric circuits.

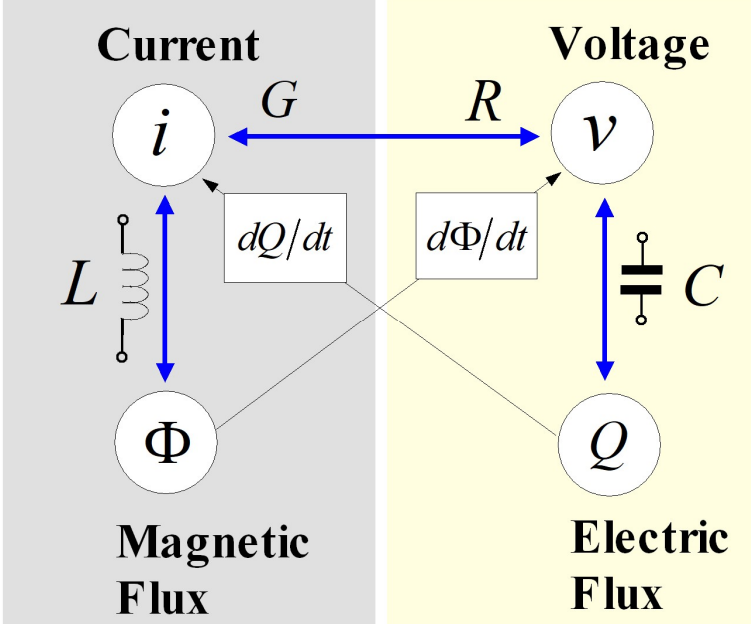

Fig. 2. The electric variables and the basic elements used in the conventional circuit analysis, where the current $i$ is the derivative of magnetic flux $\Phi$, and the voltage $v$ is the derivative of electric charge $Q$; $R$ is the resistance; $G = 1/R$ is the conductance; $L$ is the inductance; $C$ is the capacitance.

In a loop as shown in Fig. 3(a), the flux of the magnetic $B$-field thrusting through the surface of the loop is defined as $\Phi_S$. Similarly, according to the Gauss theorem, the flux of the electric $E$-field penetrating through the surface of a node is proportional to the electric charge $Q_S$ stored in the node, as shown in Fig. 3(c). Their definitions are

$$\begin{cases} \Phi_S \equiv \int_S B \cdot dS \\ Q_S \equiv \oint_S \varepsilon E \cdot dS \end{cases} \quad (1)$$

Here, $\varepsilon$ is the dielectric constant.

Therefore, to keep the consistency with the definition of the magnetic flux ($\Phi$) in the loop, we redefine the variable of electric charge ($Q$) as the electric flux enclosed in the node.

From the perspective of the electromagnetic energy transmission, the electric circuit is a system that transforms the electromagnetic energy through the electric and magnetic fluxes among nodes and loops. The dynamics of magnetic and electric flux distributions inside the circuit network can be depicted with the circuit equations and dynamic models.

Fig. 3. The basic circuit entities driven by the electric field and the magnetic field: (a) a node stores the electric-fluxes, where "+" denotes the positive free charges; (b) a non-capacitive branch Branch-$m$ with the current $i_{Bn}$ flowing inside; (c) a circuit loop contains the magnetic-fluxes, where "×" denotes the magnetic-field penetrating into the loop surface; (d) a non-inductive branch Branch-$m$, which bears the voltage $v_{Bm}$ in the loop.

### B. Electric- and Magnetic-Flux-Throughput of Branches

From the view of the electric field, the electric fluxes are stored in the capacitance network interconnected between nodes; their distributions are altered by the branches consists of non-capacitive elements.

If the branch shown in Fig. 3(b) is one of the non-capacitive branches connected to the node shown in Fig. 3(a), it pumps the electric charges to the node with a current $i_B$. According to the current-charge relation, we introduce a nominal electric flux $Q_B$ to describe the electric charge contribution of the non-capacitive branch to the node, i.e.,

$$i_B = \frac{dQ_B}{dt} \quad (2)$$

Based on this concept, the non-capacitive branch to the node works like the electric flux pump, which transfers the electric charges from one node to another.

Similarly, from the view of the magnetic field, the magnetic fluxes are preserved by the self-inductance of loop and the mutual-inductance between loops; their distributions are modified by the branches composed of non-inductive elements.

If the branch shown in Fig. 3(d) is one of the non-inductive branches inside the loop shown in Fig. 3(c), Its two terminals bear a voltage $v_B$ from the loop. According to the voltage-flux relation shown in Fig. 2, we define a nominal magnetic flux $\Phi_B$ to express the magnetic flux contribution of non-inductive branch to the loop, i.e.,

$$v_B = \frac{d\Phi_B}{dt} \tag{3}$$

In this way, the non-inductive branch inside the loop works like the magnetic flux pump, which transfers the magnetic fluxes from loop to loop.

*C. Conservation Laws of Electric- and Magnetic-Flux*

For a node connected with $Z$ non-capacitive branches, we can transform the Kirchhoff's Current Law (KCL) into the electric-flux conservation law as

$$\sum_{j=1}^{Z} i_{Bj} = \frac{dQ_S}{dt} \Rightarrow \sum_{j=1}^{Z} Q_{Bj} = Q_S + Const \tag{4}$$

It states that the $Q_S$ stored in the node equals to the algebraic sum of the $Q_{Bj}$ generated by the electric flux pumps connected to the node.

Meanwhile, for a loop consisted of $Z$ non-inductive branches, we can draw the magnetic-flux conservation law from the Kirchhoff's Voltage Law (KVL) as

$$\sum_{j=1}^{Z} v_{Bj} = \frac{d\Phi_S}{dt} \Rightarrow \sum_{j=1}^{Z} \Phi_{Bj} = \Phi_S + Const \tag{5}$$

It shows that the $\Phi_S$ preserved in the loop equals to the algebraic sum of $\Phi_{Bj}$ contributed by the magnetic flux pumps interconnected inside the loop.

If *Const* =0 in (4) and (5), two circuit laws are suitable for both superconducting and non-superconducting circuits

*D. Branches Transporting Electric- and Magnetic-Flux*

With the concepts of the nominal electric and magnetic fluxes, the current-voltage function of the non-inductive branch that works as the electric-flux pump to the node will be rewritten with the voltage-to-electric-flux function; the current-voltage function of the branch which works as the magnetic flux pump in the loop will be redefined with the current-to-magnetic-flux function.

Josephson junction and the QPS junction are two quantum elements in superconducting circuits, as shown in Fig. 4. According to Josephson equations [15], if we treat the Josephson junction shown in Fig. 4(a) as a magnetic flux pump. the Josephson current $i_J$ and the magnetic flux contribution $\Phi_J$ achieve the function $i_J = I_0 \sin(2\pi\Phi_J/\Phi_0)$, where $I_0$ is the critical current of Josephson junction, $\Phi_0 = h/2e = 2.07\times10^{-15}$ Wb.

The QPS junction shown in Fig. 4 (c) is dual to the Josephson junction according to the QPS theory [6]. Therefore, it is treated as the nonlinear electric flux pump; the relation of the voltage $v_J$ at two terminals and the electric-flux $Q_J$ is written as $v_J = V_0 \sin(2\pi Q_J/Q_0)$, where $V_0$ is the critical voltage and $Q_0 = 2e$ is the electric-flux quanta.

Fig. 4 (b) depicts the practical Josephson junction based on the Resistively-Capacitively-Shunted-Junction (RCSJ) model [16]. Using the concept of the nominal magnetic flux, its current-to-magnetic-flux function is expressed as

$$\begin{aligned} i_B &= I_0 \sin(2\pi\Phi_B/\Phi_0) + G_B\frac{d\Phi_B}{dt} + C_B\frac{d^2\Phi_B}{dt^2} + I_n(t) \\ &\equiv J_L(I_0, G_B, C_B, \Phi_B) \end{aligned} \tag{6}$$

Where $G_B$ is the conductance of the junction shunt resistor; $I_n(t)$ is the thermal current noise inside the junction.

The practical QPS junction dual to the RCSJ model is shown in Fig. 4(d). It consists of the ideal QPS junction in series with an inductor $L_B$ and resistor $R_B$; its voltage-to-electric-flux function is accordingly written as

$$\begin{aligned} v_B &= V_0 \sin(2\pi Q_B/Q_0) + R_B\frac{dQ_B}{dt} + L_B\frac{d^2Q_B}{dt^2} + V_n(t) \\ &\equiv J_C(V_0, R_B, L_B, Q_B) \end{aligned} \tag{7}$$

The definitions of Josephson junction and QPS junction show that, it is better to treat the Josephson junction circuit as the magnetic flux distribution network interconnected with the magnetic flux pumps, while it is better to treat the QPS junction circuit as the electric flux distribution network interconnected with the electric flux pumps.

If $I_0 = 0$ in (4) and $V_0 = 0$ in (5), the branch shown in Fig. 4(b) is degraded into a normal non-inductive branch with resistor and capacitor in parallel; the branch shown in Fig. 4(d) is a normal non-capacitive branch with inductor and resistor in series. Thus, the functions defined in (4) and (5) are general for both superconducting and non-superconducting circuits.

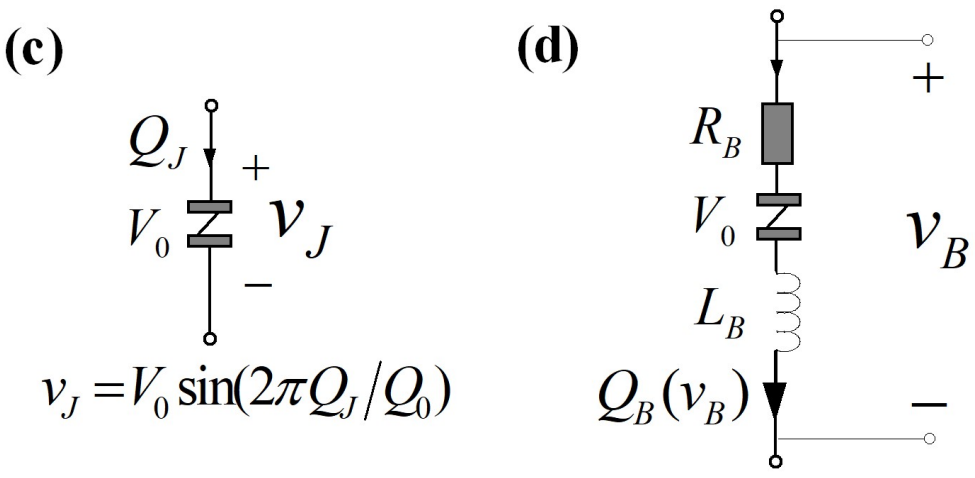

Fig. 4. The superconducting quantum elements and the practical junctions: (a) the ideal Josephson junction; (b) the practical Josephson junction; (c) the ideal QPS junction; (d) the practical QPS junction.

*E. Magnetic-Flux-Distribution Network*

For the magnetic-flux distribution circuit, its circuit topology is described with a set of loops, as shown in Fig. 5. There are ($P$+$W$-1) independent loops [12], each loop contains a mesh current which is the inner current circulating along the branches of loop. Those loops are magnetically coupled through their mutual inductances, and are electrically coupled by sharing the non-inductive branches.

In Fig. 5, those independent loops are divided into two groups, according to the value of mesh currents. The first group contains $P$ loops, each loop is defined with a mesh current to be solved, namely $i_{Mj}$ ($j$ =1, 2, …$P$). The second group includes $W$ loops, each of which is driven by the externally applied current sources $I_{bj}$ ($j$ =1, 2, …$W$-1).

In particular, the last loop, namely Loop-($P$+$W$), is defined as the outer-loop, in which the mesh current $i_{M(P+W)}$ is fixed as zero, i.e., $i_{M(P+W)} = I_{bW}$ =0. It consists of the branches that appear only once in loops from Loop-1 to Loop-($P$+$W$-1). The outer-loop makes sure that every branch is shared at least by two loops.

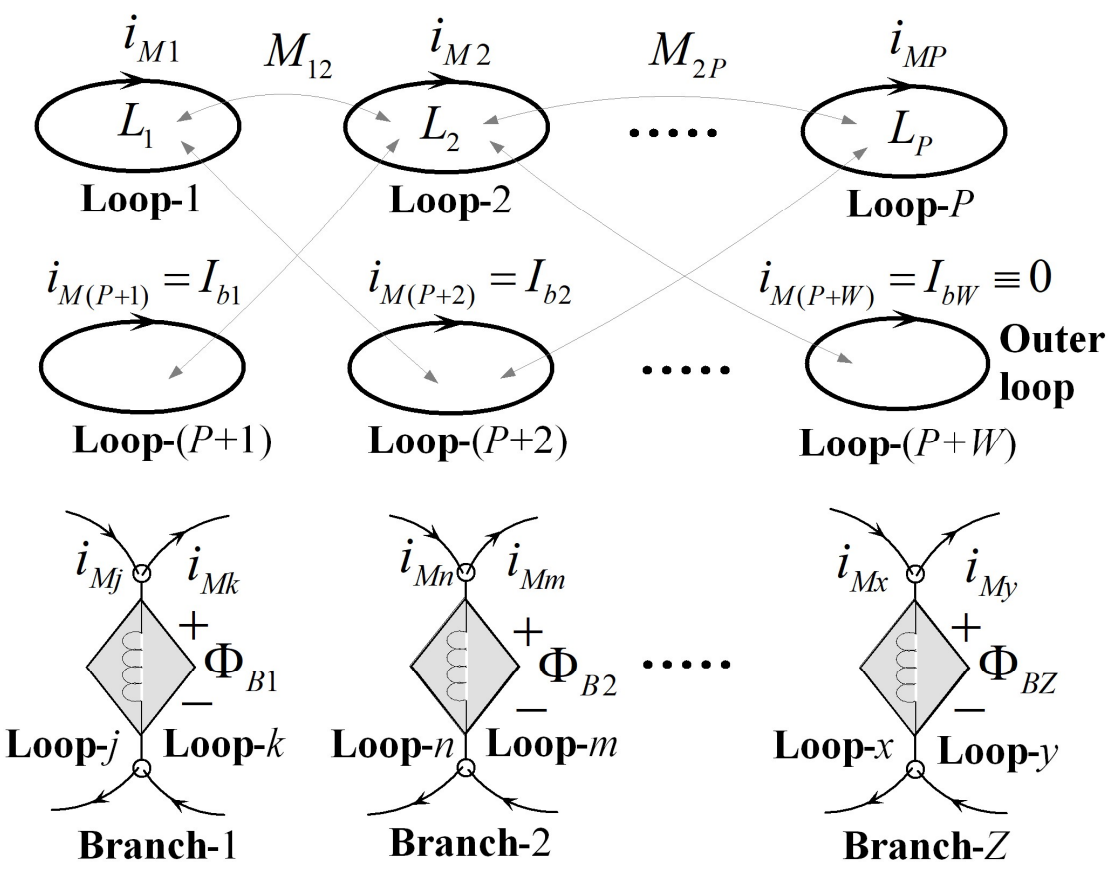

Fig. 5. A circuit composed of a group of inductively coupled loops and the non-inductive branches

The circuit analysis focus on the first $P$ loops; they are interconnected by $Z$ non-inductive branches. The magnetic fluxes coupled in the first $P$ loops are defined as $[\Phi_S] = [\Phi_{S1}, \ldots, \Phi_{SP}]$. The magnetic flux contributions of $Z$ non-inductive branches are defined as $[\Phi_B] = [\Phi_{B1}, \ldots, \Phi_{BZ}]$.

Because the mesh currents flow into and out of nodes without electric charge increment, the electric flux conservation law is always satisfied in nodes. Thus, the relation between $[\Phi_S]$ and $[\Phi_B]$ in loops can be derived by using the magnetic-flux conservation law defined in (5).

The magnetic flux conservation law implemented in $P$ loops is expressed as

$$[\sigma_M][\Phi_B]^T = [\Phi_S]^T = [L][i_M]^T \tag{8}$$

where the matrix $[\sigma_M]$ describes how those branches are connected inside $P$ loops, i.e.,

$$[\sigma_M] = \begin{matrix} & \begin{matrix} Branch{-}1 & \cdots & Branch{-}Z \end{matrix} \\ \begin{matrix} Loop{-}1 \\ \vdots \\ Loop{-}P \end{matrix} & \begin{bmatrix} \sigma_{M11} & \cdots & \sigma_{M1Z} \\ \vdots & \cdots & \vdots \\ \sigma_{MP1} & \cdots & \sigma_{MPZ} \end{bmatrix} \end{matrix} \tag{9}$$

The element $\sigma_{Mij}$ describes the polarity of $\Phi_{Bj}$ in Loop-$i$, i.e.,

$$\sigma_{Mij} = \begin{cases} +1, \text{ if } i_{Mi} \text{ flows into "}-\text{" of Branch-}j \\ -1, \text{ if } i_{Mi} \text{ flows into "}+\text{" of Branch-}j \\ 0, \quad \text{otherwise} \end{cases} \tag{10}$$

Here, the "+" and "-" terminals of Branch-$j$ are defined as shown in Fig. 4.

Meanwhile, the flux $[\Phi_S]$ distributed in $P$ loops is contributed by all the mesh currents $[i_M] = [i_{M1}, \ldots, i_{M(P+W-1)}]$ through the self-inductances and mutual-inductances. The magnetic coupling between loops is described with the matrix $[L]$ as

$$[L] = \begin{matrix} & \begin{matrix} Loop{-}1 & \cdots & Loop{-}(P+W-1) \end{matrix} \\ \begin{matrix} Loop{-}1 \\ \vdots \\ Loop{-}P \end{matrix} & \begin{bmatrix} l_{11} & \cdots & l_{1(P+W-1)} \\ \vdots & \cdots & \vdots \\ l_{P1} & \cdots & l_{P(P+W-1)} \end{bmatrix} \end{matrix} \tag{11}$$

where the element $l_{ij}$ is defined as

$$l_{ij} = \begin{cases} L_i, & i = j \\ -M_{ij}, & i \neq j \end{cases} \tag{12}$$

here, $L_i$ is the self-inductance of Loop-$i$. $M_{ij}$ is the mutual inductance between Loop-$i$ and Loop-$j$.

The magnetic flux contribution $[\Phi_B]$ from non-inductive branches is generated by mesh currents $[i_M]$. For example, in Fig. 5, $\Phi_{B1}$ is the magnetic flux contribution of Branch-1 to Loop-$j$ and Loop-$k$, when Branch-1 is driven by the mesh current $i_{Mj}$ and $i_{Mk}$. Since the branch current $i_B$ in (4) is supplied by the mesh current $[i_M]$, the non-inductive branches implement the mesh current to magnetic flux conversion among loops. Thus, the current-to-magnetic-flux function in (4) of Branch-$j$ ($j$ =1, … $Z$) can be expressed in a general form as

$$f_{Lj}(\Phi_{Bj}, [i_M]) = 0 \tag{13}$$

By synthesizing the magnetic flux distribution function of loops and the current-to-magnetic-flux functions of branches, we can draw the dynamic model of magnetic-flux distribution circuit, as shown in Fig. 6. It is a feedback system, in which non-inductive branches pump magnetic fluxes to loops, while the inductively-coupled loops generate the mesh currents to preserve magnetic fluxes according to the magnetic flux conservation law.

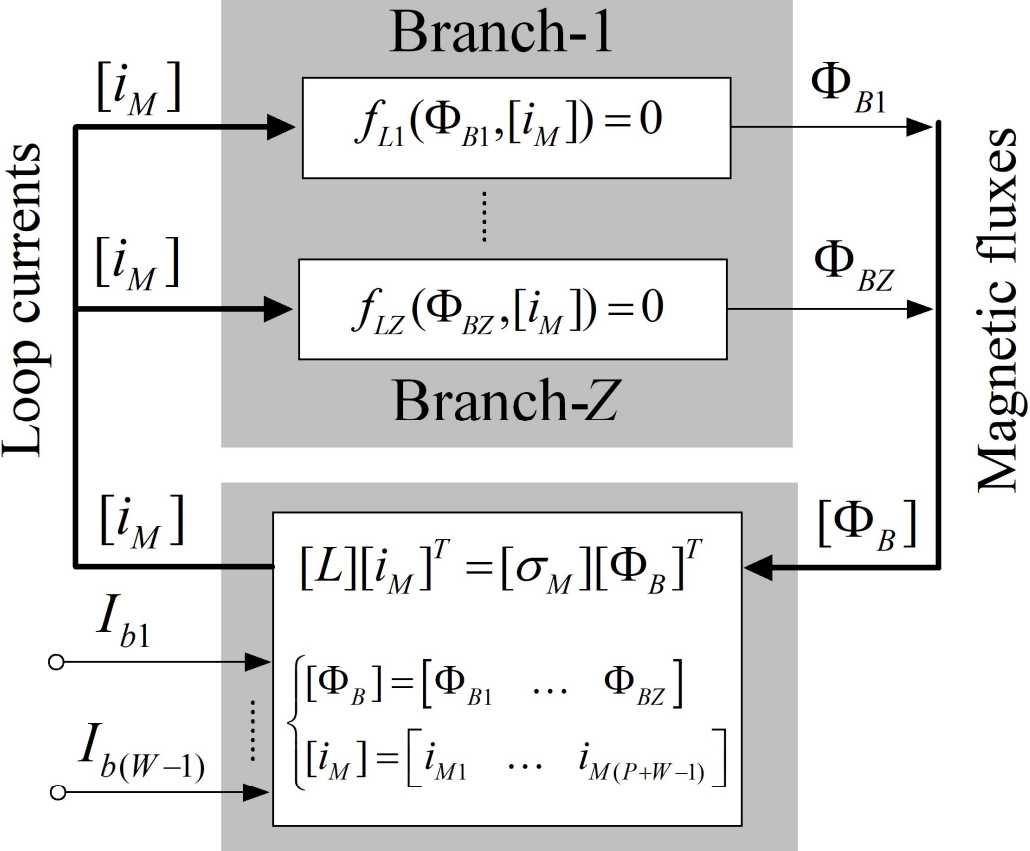

Fig. 6. The dynamic system model of the magnetic-flux distribution network.

## F. Electric-Flux-Distribution Network

If a circuit is treated as the electric flux distribution circuit, its circuit topology can be decomposed into a set of nodes interconnected with branches, as shown in Fig. 7. Assuming that there are $P$+$W$ nodes, each of which is defined with a node-voltage which is the potential relative to the ground. Those nodes are interconnected by non-inductive branches; they store electric charges through the capacitance network interconnected between them.

In Fig. 7, the voltages $v_{Nj}$ ($j$ =1, 2, …$P$) defined for the first $P$ nodes are the variables to be solved. The voltages for the other $W$ nodes are given by the external voltage source $V_{bj}$ ($j$ =1, 2, …$W$-1). Finally, the Node-($P$+$W$) is the reference node grounded at the zero voltage.

The circuit analysis focus on the first $P$ nodes; they are interconnected by $Z$ branches; those branches modify the electric flux distribution with electric flux contributions. We first define the electric fluxes contained in the node as $[Q_S] = [Q_{S1}, \ldots, Q_{SP}]$, then define the electric-flux contributions of branches as $[Q_B] = [Q_{B1}, \ldots, Q_{BZ}]$.

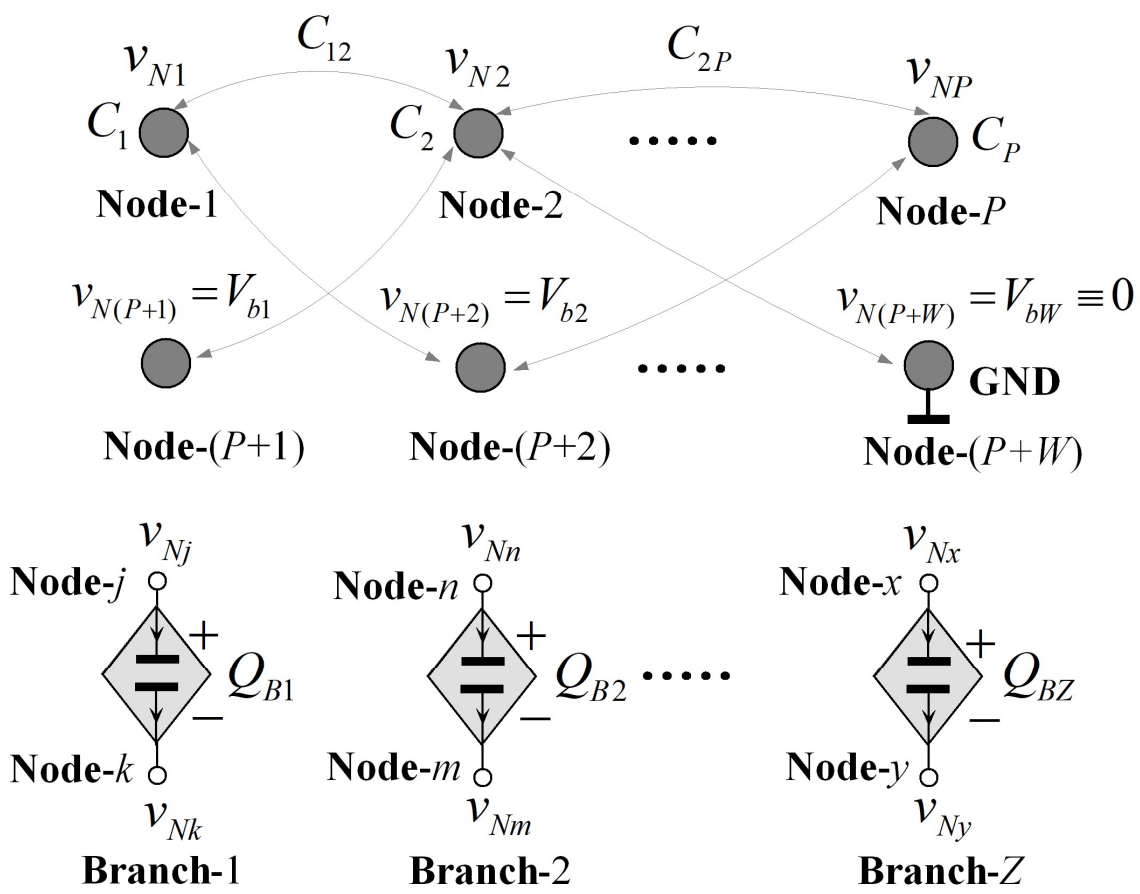

Fig. 7. A circuit network displayed with a group of capacitively-coupled nodes and the non-capacitive branches.

Because the definition of the node-voltage indicates that electric field in any loop achieves no curl, the magnetic-flux conservation law is always satisfied in the nodes of electric-flux distribution circuits. The circuit equations can be derived by implementing the electric-flux conservation law to $P$ nodes.

According to the electric-flux conservation law. The electric-flux distribution function is expressed as

$$[\sigma_N][Q_B]^T = [Q_S]^T = [C][v_N]^T \tag{14}$$

where $[\sigma_N]$ describes how those branches are connected to the nodes. i.e.,

$$[\sigma_N] = \begin{matrix} \scriptstyle Node-1 \\ \vdots \\ \scriptstyle Node-P \end{matrix} \overset{\begin{matrix} \scriptstyle Branch-1 & \cdots & \scriptstyle Branch-Z \end{matrix}}{\begin{bmatrix} \sigma_{N11} & \cdots & \sigma_{N1Z} \\ \vdots & \cdots & \vdots \\ \sigma_{NP1} & \cdots & \sigma_{NPZ} \end{bmatrix}} \tag{15}$$

The element $\sigma_{Nij}$ is defined as

$$\sigma_{Nij} = \begin{cases} +1, & \text{if } i_{Bj} \text{ flows into Node-}i \\ -1, & \text{if } i_{Bj} \text{ flows out of Node-}i \\ 0, & \text{otherwise} \end{cases} \tag{16}$$

The $[Q_S]$ are stored in $P$ nodes under the node-voltages $[v_N]$ through the capacitor network. The capacitances between nodes are described with matrix $[C]$ as

$$[C] = \begin{matrix} \scriptstyle Node-1 \\ \vdots \\ \scriptstyle Node-P \end{matrix} \overset{\begin{matrix} \scriptstyle Node-1 & \cdots & \scriptstyle Node-(P+W-1) \end{matrix}}{\begin{bmatrix} c_{11} & \cdots & c_{1(P+W-1)} \\ \vdots & \cdots & \vdots \\ c_{P1} & \cdots & c_{P(P+W-1)} \end{bmatrix}} \tag{17}$$

Where the element $c_{ij}$ is defined as

$$c_{ij} = \begin{cases} C_i, & i = j \\ -C_{ij}, & i \neq j \end{cases} \tag{18}$$

Here, $C_i$ is the total capacitance of Node-$i$. $C_{ij}$ is the mutual capacitance between Node-$i$ and Node-$j$.

The $[Q_B]$ is contributed by the non-capacitive branches under the node-voltages $[v_N]$. For example, Branch-1 shown in Fig. 7 is connected between Node-$j$ and Node-$k$; it transfers $\Phi_{B1}$ from Node-$j$ to Node-$k$. The voltage $v_B$ at two terminals of the branch can be expressed with $[v_N]$. Thus, the current-to-magnetic-flux function for the non-capacitive Branch-$j$ ($j$ =1, 2, … $Z$) can be expressed in a general form as

$$f_{Cj}(Q_{Bj},[v_N]) = 0 \tag{19}$$

Similarly, by synthesizing all the circuit equations, the dynamic system model of the electric flux distribution circuit is depicted in Fig. 8, it achieves the same dynamic working principles with the system model shown in Fig. 6.

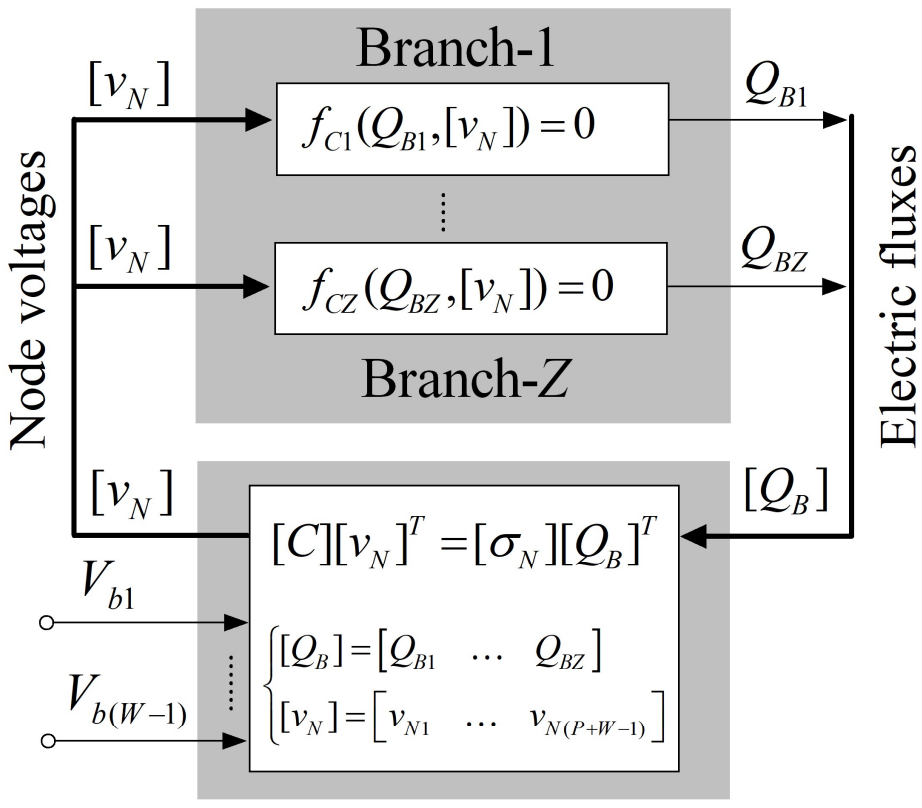

Fig. 8. The dynamic system model of the electric-flux distribution network.

## G. *Duality Principle*

Fig. 6 exhibits the mathematic relation between $[\Phi_B]$ and $[i_M]$ for the magnetic-flux distribution circuit; Fig. 8 depicts the mathematic relation between $[Q_B]$ and $[v_N]$ for the electric-flux distribution circuit. Circuit equations shown in Fig. 6 and Fig. 8 are in duality. The duality principle between two circuits is summarized in Table I.

TABLE I. THE DUALITY BETWEEN CIRCUITS VIEWED BY THE ELECTRIC AND MAGNETIC FIELDS

| | Magnetic-flux distribution network | Electric-field distribution network |
| --- | --- | --- |
| Topology description | Inductively-coupled Loops | Capacitively-coupled Nodes |
| Physic quantity | Magnetic flux | Electric flux |
| Circuit variable | Loop-current | Node-voltage |
| Branch model | Magnetic-flux pump | Electric-flux pump |
| Circuit laws | Magnetic-flux conservation law | Electric-flux conservation law |

Two circuit are equivalent, if their circuit equations are in dual relations. To find an electric-flux distribution circuit dual to a magnetic-flux distribution circuit, we first define the variables transformation relations as

$$\begin{cases} [Q_B] \Leftrightarrow \dfrac{1}{r_0}[\Phi_B] \\ [v_N] \Leftrightarrow r_1[i_M] \end{cases} \tag{20}$$

Here, $r_0$ is the factor for the transformation between $[\Phi_B]$ and $[Q_B]$; $r_1$ is the transformation factor between and $[i_M]$ and $[v_N]$. According the duality relation between (6) and (7), we can find that $r_0 = \Phi_0/Q_0$ and $r_1 = V_0/I_0$ for the duality principle between Josephson junction and QPS junction circuits.

To keep two circuits equivalent, their circuit parameters will meet the condition as

$$\begin{cases} [\sigma_N]=[\sigma_M];[C]=\dfrac{[L]}{r_0 r_1} \\ f_{Cj}(Q_{Bj},[v_N]) = f_{Lj}(r_0 Q_{Bj}, \dfrac{[v_N]}{r_1}),(j=1,\cdots,Z) \end{cases} \tag{21}$$

Because a two-terminal branch can be shared by more than two loops in the magnetic-flux distribution circuit, but can only be connected between two nodes in the electric-flux distribution circuit, the condition $[\sigma_M]$ = $[\sigma_N]$ can only be satisfied for those magnetic-flux distribution circuits in which the branches are inserted between two plane loops. Therefore, a given QPS junction circuit may find an equivalent Josephson junction circuit, but not all the Josephson junction circuits have an equivalent QPS junction circuit, although we can always find a non-capacitive branch with QPS junction dual to a given non-inductive branch with Josephson junction according to the topologies shown in Fig. 4(b) and (d).

## III. An Example

The same mathematic structure shown in Fig. 6 and Fig. 8 exhibits the dynamics of electric circuit from the views of magnetic and electric fields respectively. We call it the electromagnetic-flux-distribution model.

This model provides a quick approach to derive the circuit equation for both Josephson junction and QPS junction circuits. The circuit equation group includes a linear matrix equation and the functions of $Z$ branches. The simulation for both Josephson junction and QPS junction circuits is simply implemented by finding the numerical solutions of the second-order differential equations of branches.

This section will illustrate the application of the electromagnetic-flux distribution model through analyses of the SQUID circuit shown in Fig.1 and its dual circuit.

The dc SQUID circuit shown in Fig. 1 is redrawn with an equivalent circuit, as shown in Fig. 9 (a), where the external flux $\Phi_e$ applied to the dc SQUID is implemented with a loop driven by the current source $I_e$.

It is a magnetic-flux distribution circuit with four loop currents $[i_M]$ = [ $i_{M1}$, $i_{M2}$, $i_{M3}$, $i_{M4}$], where the mesh-current $i_{M3}$ and $i_{M4}$ are supplied by current sources $I_e$ and $I_b$. There are three non-inductive components with magnetic-flux contributions $[\Phi_B]$ = [ $\Phi_{B1}$, $\Phi_{B2}$, $\Phi_{B3}$], where $\Phi_{B1}$ and $\Phi_{B2}$ are the magnetic-flux contributions of $J_1$ and $J_2$; $\Phi_{B3}$ is for the branch of $R_O$ and $C_O$ in parallel.

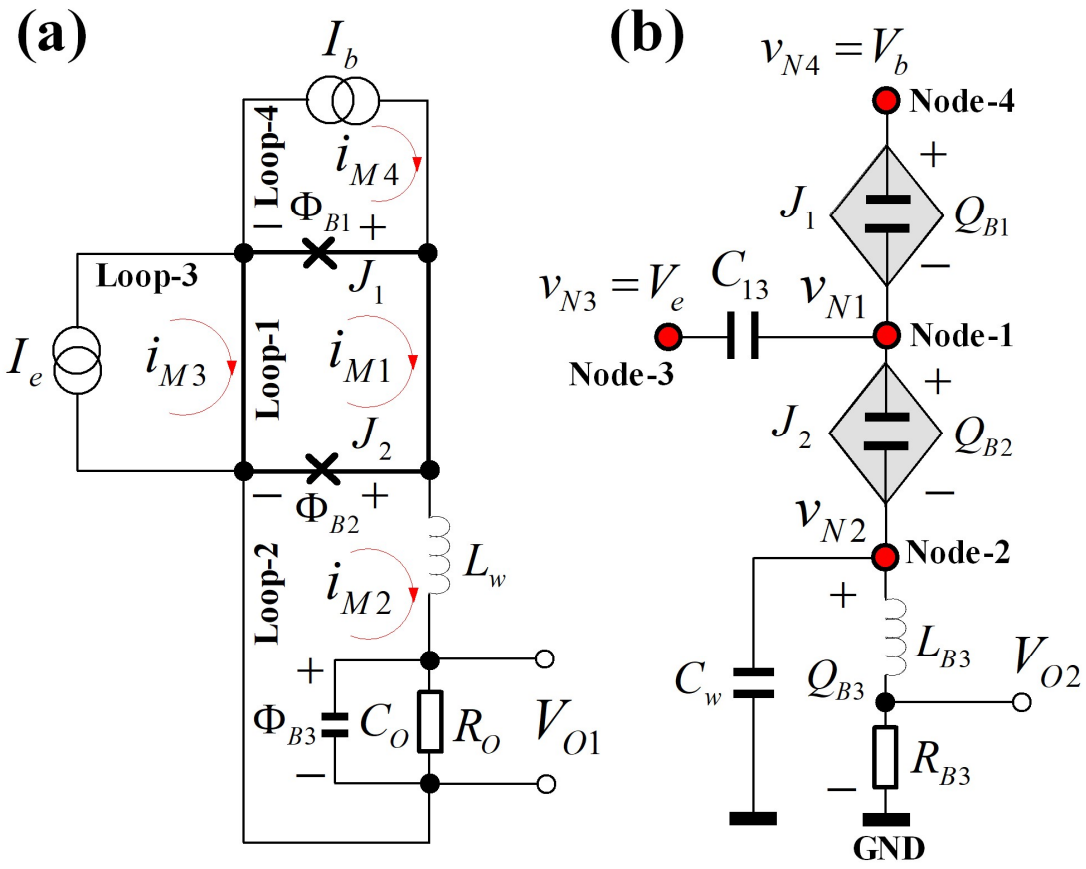

Fig. 9. Examples of dc SQUID and its dual circuits: (a) the dc SQUID circuit spread out with loops; (b) the dual circuit based on the QPS junctions.

The dynamics model of the SQUID circuit is depicted in Fig. 10. Three magnetic-flux-pumps driven by the mesh currents $[i_M]$ generate the magnetic-fluxes $[\Phi_B]$ to the loops; the loop-currents $[i_M]$ are changing with the $[\Phi_B]$ in loops according the linear matrix equations derived in (8).

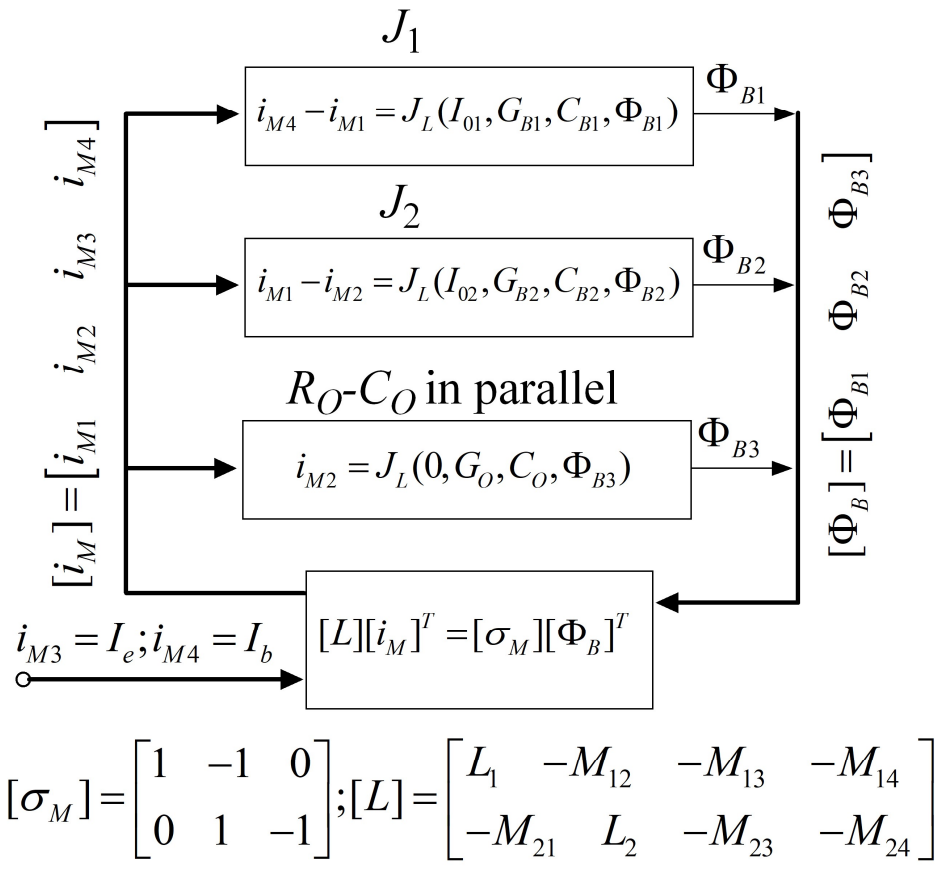

Fig. 10. The dynamic system model of the given dc SQUID circuit.

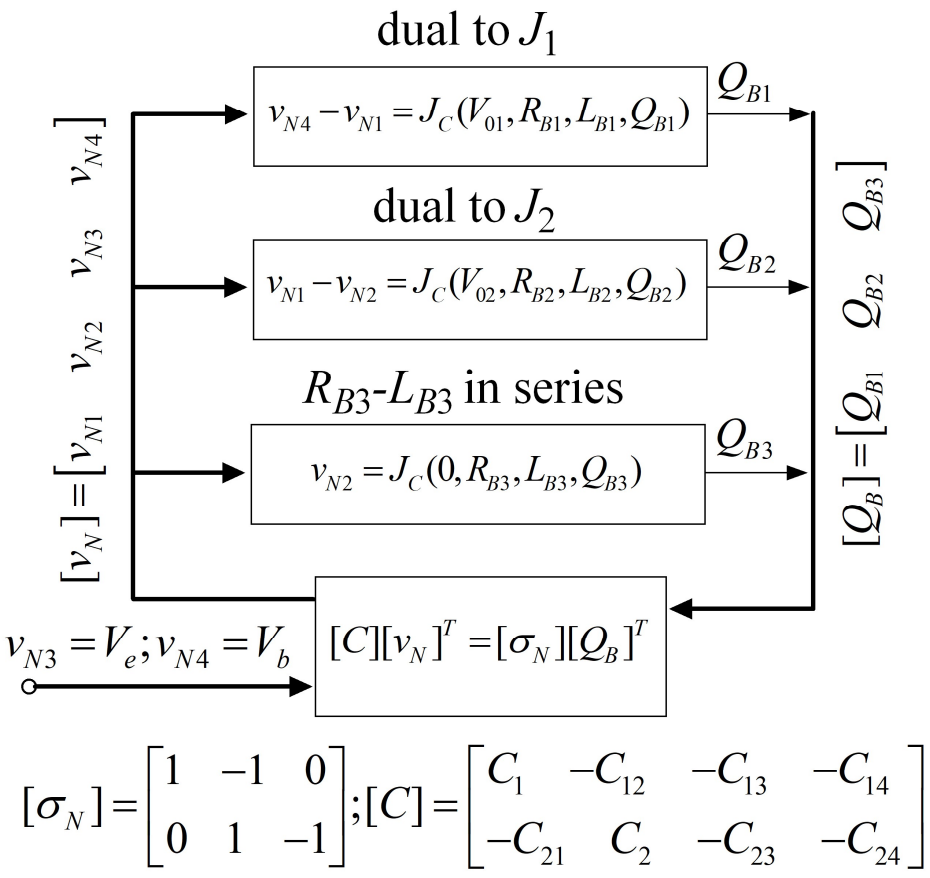

Fig. 11. The dynamic system model of the QPS junction circuit dual to the given dc SQUID circuit.

Fig. 11 shows the electric-flux-distribution model transformed from Fig. 10 according to the duality principles. In this model, three electric-flux-pumps driven by node-voltage $[v_N]$ generate the electric-flux $[Q_B]$ to the nodes; while the node-voltage $[v_N]$ is varied by the electric-flux $[Q_B]$ according to the linear matrix equations derived in (14).

This electric-flux-distribution circuit implemented with QPS junctions is shown in Fig. 9(b). It is the equivalent circuit dual to the Josephson junction circuit shown in Fig. 9(a). In Fig. 9(a), $V_{O1}$ is the average voltage measured at the output of the SQUID circuit, i.e., $V_{O1} = R_O < i_{M2} >$, while $V_{O2}$ is the voltage output of the QPS circuit, $V_{O2} = < v_{N2} >$. Here, $< i_{M2} >$ and $< v_{N2} >$ are the average values of $i_{M2}$ and $v_{N2}$.

The FET [18] shown in Fig. 1 is an electric filed sensitive device, and works for the electric-flux distribution rather than the magnetic-flux distribution. Therefore, the SQUID and FET hybrid circuit contains both the magnetic-flux distribution and the electric-flux distribution networks. Its analysis can be unified, if we replace the SQUID circuit with its equivalent QPS junction circuit shown in Fig. 9(b).

## IV. Simulations

### A. Simulation Results

The dynamic models in Fig. 10 and Fig. 11 are simulated by solving the same mathematic equations. The parameters for SQUID circuit simulation are shown in Table II. Here, the capacitance of Josephson junctions is expressed with the Stewart-McCumber parameter [17].

TABLE II. Component Parameters for Simulations

| Parameter | Symbol | Value(Unit) |
|---|---|---|
| The critical current of $J_1$ and $J_2$ | $I_{01}$, $I_{02}$ | 10 μA |
| Shunt resistor of $J_1$ and $J_2$ | $R_1$, $R_2$ | 6.3 Ω |
| Stewart-McCumber Parameter | $\beta_{C1}$, $\beta_{C2}$ | 0.35 |
| Loop inductance of dc SQUID | $L_s$ | 60 pH |
| Mutual inductances | $M_{12}$, $M_{13}$, $M_{14}$ | 20 pH |
| The inductance of inductor $L_a$ | $L_w$ | 100 nH |
| Resistor | $R_O$ | 10 kΩ |
| Capacitor | $C_O$ | 10 pF |
| Mutual inductances | $M_{23}$, $M_{24}$ | 0 pH |

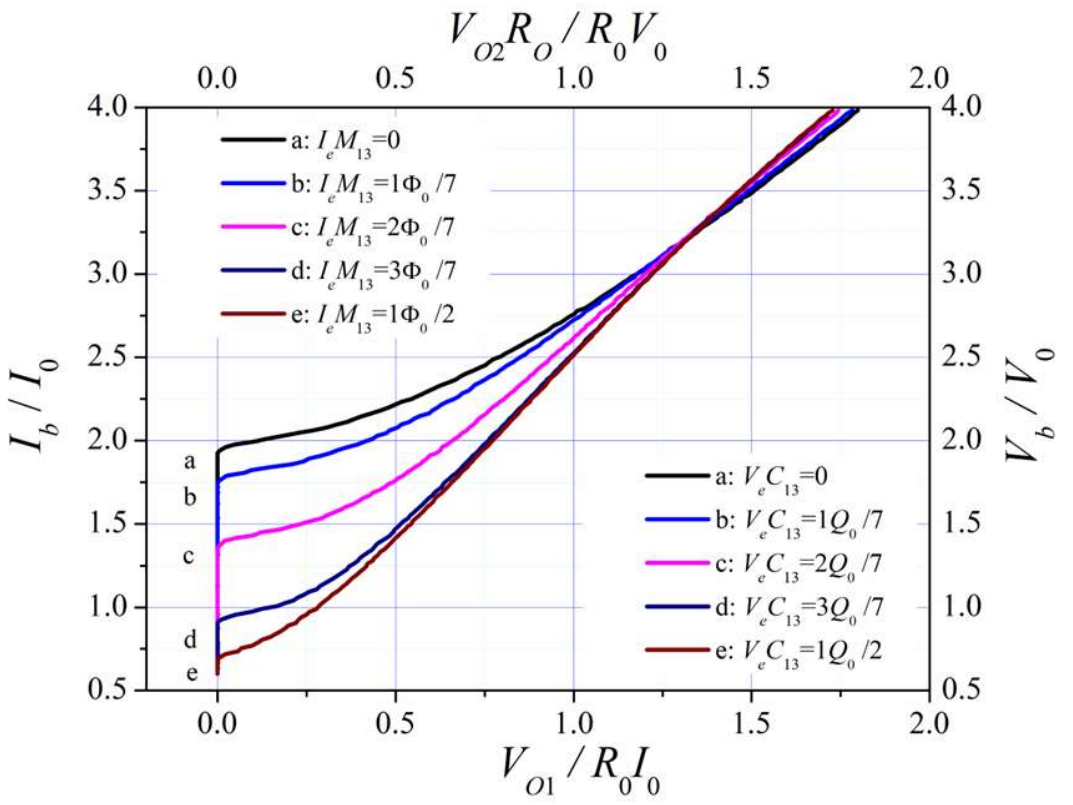

Fig. 12. The magnetic-flux modulated current-to-voltage characteristics of the dc SQUID circuit, and the electric-flux modulated voltage-to-voltage characteristics of the QPS junction circuit.

Fig. 12 depicts the current-voltage characteristics of the Josephson junction circuit, as well as the voltage-voltage characteristics of the equivalent QPS junction circuit. It shows that $V_{O1} / V_{O2} = R_O / r_1$, thus, the outputs of two circuits are different unless $R_O = r_1$.

Fig. 13 exhibits the periodical flux-voltage characteristics of the Josephson junction circuit and the periodical charge-voltage characteristic of the QPS junction circuit. This feature is introduced by the sine function in (6) and (7); the periods of the characteristics are quantized as $\Phi_0$ and $Q_0$ respectively.

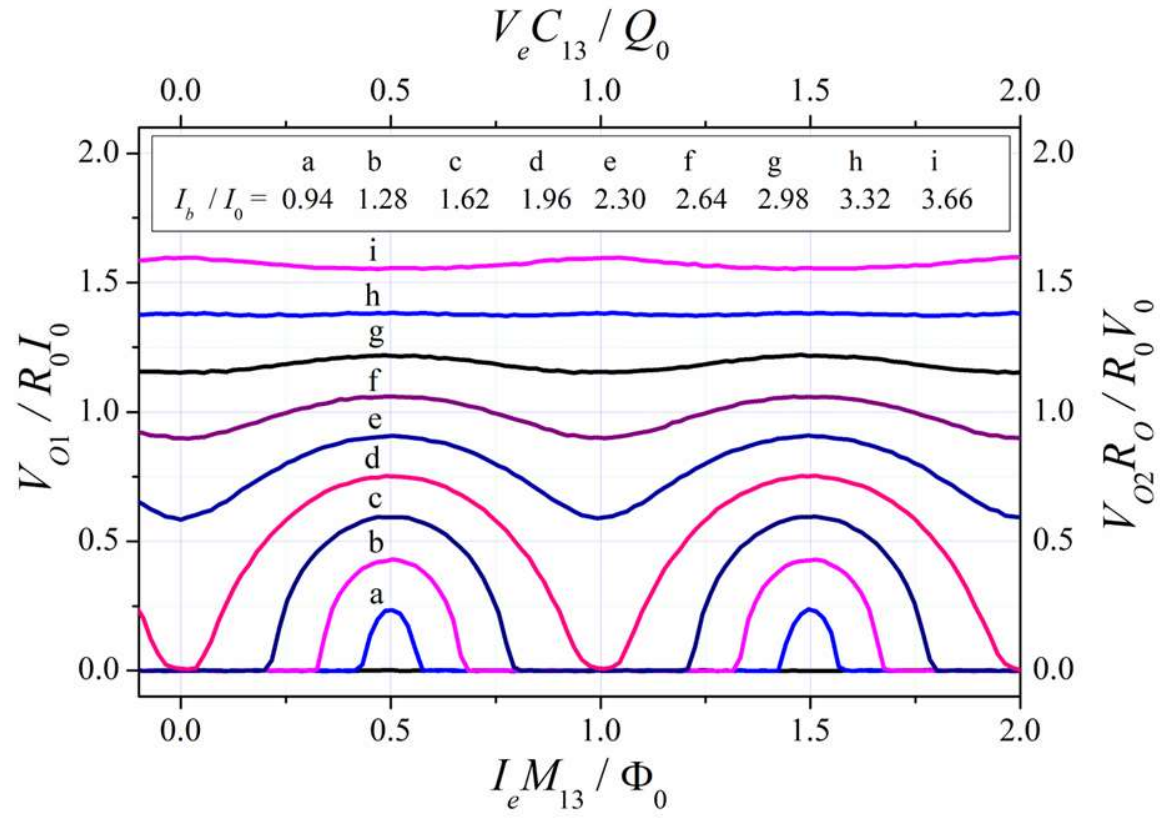

Fig. 13. The periodical magnetic-flux-to-voltage characteristics of the dc SQUID circuit and the electric-flux-to-voltage characteristics of the QPS junction circuit.

### B. Discussion

The electromagnetic-flux-distribution model achieves four advantages in the analyses of Josephson junction and QPS junction circuits.

First, this model uses the electric or magnetic flux instead of the quantum phase to define the functions of Josephson junctions and QPS junctions. Those circuit concepts and circuit element definitions are easily understood by the electronic engineers who are unfamiliar with the superconducting physics.

Second, this model drives the unified circuit equations for both the Josephson junction circuits and QPS junction circuits. With only the circuit parameters of network and branches, the unified circuit equation simplifies the circuit analysis and simulation of superconducting circuits significantly.

Third, the electromagnetic-flux-distribution model depicts the dynamics of circuit elements working in the electromagnetic fields. In Josephson junction circuits, the non-inductive branches with Josephson junction work as the magnetic-flux pumps, and are interfering each other inside loops, according to the magnetic-flux conservation law. Similarly, in QPS junction circuits, the non-capacitive branches with QPS junction work as the electric-flux pumps, and are interfering each other in nodes, according to the magnetic-flux conservation law.

Finally, the electromagnetic-flux-distribution model derives the duality principle between the and the QPS junction circuits. The mathematically expressed duality principle enables us to find an equivalent Josephson junction circuit for any QPS junction circuits precisely.

Both the Josephson junction circuits and the QPS junction circuits will become a normal circuit consisted of the *RLC* elements, if $V_0$=0 in (6) and $I_0$= 0 in (7). It indicates that, the superconducting Josephson junction circuits and QPS junction circuits are the normal *RLC* circuit modulated by the additional Josephson junctions and QPS junctions.

The electromagnetic-flux-distribution model unifies both the superconducting and the non-superconducting circuits.

## V. Conclusion

We introduce a general electromagnetic-flux-distribution model to unify the circuit analysis and simulation for both the superconducting and non-superconducting circuits. Compared to the conventional non-superconducting circuit, the unique element of the superconducting circuits is the sine function introduced by the Josephson junction and the QPS junction. A Josephson junction circuit is a magnetic-flux-distribution system, in which the branches with Josephson junctions work as the magnetic-flux pumps, and are exchanging the magnetic fluxes inside loops according to the magnetic-flux conservation law. Meanwhile, a QPS junction circuit functions as an electric-flux-distribution system, in which the branches with QPS junctions work as the electric-flux pumps, and are exchanging the electric fluxes in nodes according to the electric-flux conservation law. It is shown that Josephson junction and the QPS junction circuits are in duality by implementing the same dynamic model through different electromagnetic fluxes.

This general electromagnetic-flux-distribution model is suitable for the superconducting-semiconducting hybrid circuit design to unify the circuit analyses of both the superconducting and non-superconducting circuits.

## References

[1] J. Clarke and A. I. Braginski, *The SQUID Handbook: Fundamentals and Technology of SQUIDs and SQUID Systems Vol. I.* New York: Wiley, 2004, pp. 1-127.

[2] R. P. Welty and J. M. Martinis, "A series array of DC SQUIDs," *IEEE Transactions on Magnetics,* vol. 27, no. 2, pp. 2924-2926, 1991.

[3] J. Oppenlander, T. Trauble, C. Haussler, and N. Schopohl, "Superconducting multiple loop quantum interferometers," *IEEE Transactions on Applied Superconductivity,* vol. 11, no. 1, pp. 1271-1274, Mar 2001.

[4] K. K. Likharev and V. K. Semenov, "RSFQ logic/memory family: a new Josephson-junction technology for sub-terahertz-clock-frequency digital systems," *IEEE Transactions on Applied Superconductivity,* vol. 1, no. 1, pp. 3-28, 1991.

[5] R. Kleiner, R. Hott, T. Wolf, G. Zwicknagl, and F. Sirois, *Applied Superconductivity : Handbook on Devices and Applications*. Weinheim,Germany: Wiley-VCH Verlag GmbH & Co., 2015, pp. 949-1110.

[6] J. E. Mooij and Y. V. Nazarov, "Superconducting nanowires as quantum phase-slip junctions," *Nature Physics,* vol. 2, no. 3, pp. 169-172, Mar 2006.

[7] T. T. Hongisto and A. B. Zorin, "Single-Charge Transistor Based on the Charge-Phase Duality of a Superconducting Nanowire Circuit," *Physical Review Letters,* vol. 108, no. 9, Feb 27 2012.

[8] A. J. Kerman, "Flux-charge duality and topological quantum phase fluctuations in quasi-one-dimensional superconductors," *New Journal of Physics,* vol. 15, Oct 18 2013.

[9] T. Van Duzer, Y. J. Feng, X. F. Meng, S. R. Whiteley, and N. Yoshikawa, "Hybrid Josephson-CMOS memory: a solution for the Josephson memory problem," *Superconductor Science & Technology,* vol. 15, no. 12, pp. 1669-1674, Dec 2002.

[10] K. K. Likharev, "Single-Electron Transistors - Electrostatic Analogs of the Dc Squids," *IEEE Transactions on Magnetics,* vol. 23, no. 2, pp. 1142-1145, Mar 1987.

[11] S. V. Polonsky, V. K. Semenov, and P. N. Shevchenko, "PSCAN: personal superconductor circuit analyser," *Superconductor Science and Technology,* vol. 4, no. 11, pp. 667-670, 1991.

[12] C. K. Alexander, M. N. Sadiku, and M. Sadiku, *Fundamentals of electric circuits*. McGraw-Hill New York, 2009.

[13] A. Davidson and M. R. Beasley, "Duality between Superconducting and Semiconducting Electronics,", *IEEE Journal of Solid-State Circuits,* vol. 14, no. 4, pp. 758-762, 1979.

[14] A. M. Kadin, "Duality and fluxonics in superconducting devices," *Journal of Applied Physics,* vol. 68, no. 11, pp. 5741-5749, 1990.

[15] T. Michael, *Introduction to Superconductivity: Second Edition*. New York: McGraw-Hill, Inc, 1996, pp. 110-130.

[16] D. E. McCumber, "Effect of ac Impedance on dc Voltage-Current Characteristics of Superconductor Weak-Link Junctions," *Journal of Applied Physics,* vol. 39, no. 7, pp. 3113-3118, 1968.

[17] W. C. Stewart, "Current-Voltage Characteristics of Josephson Junctions," *Applied Physics Letters,* vol. 12, no. 8, pp. 277-280, 1968.

[18] H. Shichman and D. A. Hodges, "Modeling and simulation of insulated-gate field-effect transistor switching circuits," *IEEE Journal of Solid-State Circuits,* vol. 3, no. 3, pp. 285-289, 1968.